\documentclass[useAMS,usenatbib]{mn2e}

\def\ergcm2s{~erg cm$^{-2}$ s$^{-1}$ } 
\def\funit{~erg cm$^{-2}$ s$^{-1}$ } 
\def\ergs{~erg s$^{-1}$}
\def\nh{$N_{H}$}

\def\etal{et al.~}

\def\msun{~M$_{\odot}$}

\def\deg{$^{\circ}$}
\def\n4038{~NGC4038/39}
\def\chandra{{\it Chandra }}

\def\int{{\it INTEGRAL }}
\def\suzaku{{\it Suzaku }}
\def\suz{{\it Suzaku }}

\def\x2{$\chi^{2}$}

\def\apj{ApJ}
\def\apjl{ApJ {\it{Lett.}}}
\def\apjs{ApJ Suppl.}
\def\aap{A\&A}
\def\pasj{PASJ}
\def\nat{Nature}
\def\mnras{MNRAS}

\usepackage{graphicx}
\usepackage{epsfig}
\usepackage{apjfonts}

\begin{document}
\title[X-ray spectra of AXPs/SGRs]{Broad-band X-ray spectra of anomalous X-ray pulsars and soft $\gamma$-ray repeaters:  pulsars in a weak-accretion regime ?}

\author[A. Zezas et al]{A. Zezas$^{1}$\thanks{E-mail: azezas@physics.uoc.gr (AZ)} J. E. Tr\"{u}mper$^{2}$\thanks{E-mail: jtrumper@mpe.mpg.de (JET)} and  N. D. Kylafis$^{1}$\thanks{E-mail: kylafis@physics.uoc.gr (NDK)} \\
$^{1}$University of Crete, Physics Department \& Institute of Theoretical \& 
Computational Physics, 71003 Heraklion, Crete, Greece\\
Foundation for Research and Technology-Hellas, 71110 Heraklion, Crete,
Greece\\
$^{2}$Max-Planck-Institut f\"{u}r extraterrestrische Physik, 
Postfach 1312, 85741 Garching, Germany }

\maketitle

\label{firstpage}

\begin{abstract}
 We present the results from the analysis of the broad-band X-ray spectra of 5 Anomalous X-ray Pulsars (AXPs) and Soft $\gamma$-ray Repeaters (SGRs). We fit their \suzaku\ and \int\ spectra with models appropriate for the X-ray emission from the accretion flow onto a pulsar. We find that their X-ray spectra can be well described with this model. In particular we find that: (a) the radius of the accretion column is $\sim150-350$\,m resulting in a transverse optical depth of $\sim 1$; (b) the vertical Thompson optical depth is $\approx 50-400$, and (c) their luminosity translates in accretion rates $\approx10^{15}\rm{g\, s^{-1}}$. These results are in good agreement with the predictions from the fall-back disk model, providing further support in the interpretation of AXPs and SGRs as accreting pulsars.   
\end{abstract}



\section{Introduction}

  The anomalous X-ray pulsars (AXPs) are objects that show X-ray pulsations but, in comparison to the more luminous accreting pulsars in X-ray binaries, they exhibit several unusual properties (e.g. no evidence for a companion star, softer X-ray spectra than accreting pulsars). Soft $\gamma$-ray Repeaters (SGRs) on the other hand are sources that produce  short ($<1$\,sec, typically $\sim0.1$\,s) bursts with luminosities up to $\sim10^{41}$\ergs in low-energy $\gamma$-rays and X-rays. They also exhibit rare, giant flares with luminosities up to  $10^{44}$\ergs,  as well as low-luminosity ($10^{35-36}$\ergs) persistent emission  (e.g. Woods \& Thompson 2006).     It is now widely accepted that SGRs and AXPs are manifestations of the same class of objects (e.g.  Kouveliotou \etal 1998; Woods \& Thompson 2006) which, in addition to the bursting activity, share some general common characteristics. These  common traits include:  low X-ray luminosities ($10^{34-36}$\ergs), but significant hard X-ray emission (e.g. Meregheti 2008; Enoto \etal 2010), relatively long periods (P$\sim1-12$s), and large spin-down rates ($\dot{P}>10^{-13}\rm{s\,s^{-1}}$) (e.g.  Woods \& Thompson 2006). 
 
 These properties have been generally interpreted within the context of the magnetar model (e.g. Duncan \& Thompson 1992; Thompson \& Duncan 1995; Beloborodov 2013). According to this model, AXPs are together with SGRs, two manifestations of non-accreting young pulsars with ultra-strong magnetic fields in the $10^{13-15}$G range  (e.g. Woods \& Thompson 2006, and references therein).  The persistent emission of magnetars is believed to be due to the decay of their strong magnetic field, while their outbursts are interpreted as violent magnetic reconnection events due to the rearrangement of plates on the crust of the pulsar.  The ultra-strong magnetic field of these systems is inferred from their spin decay. However, recently several members of the AXP/SGR family have been found to have dipole magnetic fields (as derived from their $P$ and $\dot{P}$) in the more typical range of $\sim10^{12-13}$G for accreting pulsars (e.g. Rea \etal 2014; Rea \etal 2013).  This has led to the notion that the activity of these objects is powered by an internal toroidal field, which also leads to the formation of local multipole fields.
Although the magnetar model explains the energy output  and the outbursts of AXPs and SGRs, their broad-band X-ray spectra are usually described on the basis of ad-hoc phenomenological models such as combination of  power-law components (e.g. Meregheti 2008;  den Hartog \etal 2008a, 2008b; Enoto \etal 2010). However, so far there has not been any physical model that reproduces consistently the entire broad-band X-ray spectra of AXPs/SGRs. For example, Beloborodov (2013) proposed a spectral model based on the outflow of $\rm{e^{-}-e^{+}}$ pairs produced close to the surface of the neutron star, and it was recently applied to the spectra of 3 AXPs/SGRs (Hasco{\"e}t \etal 2014), providing a good description of their hard X-ray spectra, but requiring the ad-hoc inclusion of multiple modified black-body components to fit the spectrum below 10\,keV. On the other hand the ``twisted magnetosphere'' model of Zane \etal (2009), while it fits well the spectrum below 10\,keV, it requires an additional power-law component to reproduce the hard part of the spectrum.

  An alternative model to that of an isolated spinning-down magnetar  is that of a newborn pulsar with a magnetic dipole field ($10^{11-13}$G), accreting leftover gas either from the  common-envelope phase (van Paradijs et al. 1995), or from  the fall-back disk of its parent supernova explosion (Chatterjee et al. 2000; Alpar 2001). In this case, the spin down of the neutron star is attributed to the interaction between the accretion disk and the magnetosphere (e.g. Ertan \etal 2012). Owing to the low luminosities, the accretion rate of these systems is expected to be much lower than that of typical accreting pulsars in X-ray binaries. In addition, given their low luminosity and lack of donor stars,  we would not expect to observe strong optical emission from these systems,  in agreement with the general characteristics of the AXP/SGR population. 
  In this context, small scale outbursts observed in AXPs/SGRs can be easily explained in terms of accretion of clumps in the fall-back disk. On the other hand, large outbursts are explained in the context of reconnection of local multipole magnetic fields anchored on the surface of the neutron star.     In general this ``accretion model''  reproduces well the spin period distribution, light curves, and broad-band multi-wavelength emission of AXPs and SGRs (e.g. Ertan \etal 2007; {\c C}al{\i}{\c s}kan  \& Ertan, 2012).

 The discovery of hard X-ray emission from AXPs/SGRs extending up to $>100$\,keV (e.g. Kuiper \etal 2004, 2006; den Hartog \etal 2008; Enoto \etal 2010), gives us an additional test-bed for the fall-back disk model.  
 In the first of a series of papers exploring the consistency of the hard X-ray emission from AXPs/SGRs with the ``accretion model''  (Tr\"{u}mper \etal 2010), we found that indeed the hard X-ray spectra of the brightest  AXP, 4U 0142+61, can be interpreted in terms of accretion. The \chandra and INTEGRAL spectrum of this source can be well represented by a model including thermal as well as bulk-motion Comptonization ({\tt bmc} model; e.g. Mastichiadis \& Kylafis 1992; Titarchuk \etal 2007). 
A follow-up study of the energy-dependent pulse profiles of the same source (Tr\"{u}mper \etal 2013) showed that they can be well modeled in the context of the fan beam from the accretion column, and a polar beam from the hot polar cap. 
More recently, Guo \etal (2015) analyzed a sample of 4 AXPs/SGRs with the same spectral model as in Tr\"{u}mper \etal (2010). They found that the hard X-ray spectra of these 4 additional sources can also be well reproduced with the {\tt bmc} model, further supporting the accretion model for AXPs.   
 
However, the spectral models used in the studies of Tr\"{u}mper \etal (2010) and Guo \etal (2015) have several shortcomings when applied to accreting pulsars: they assume spherical or  disk geometries, instead of the cylindrical geometry of the accretion flow onto a pulsar, and they do not account for the different scattering cross-sections in the presence of a strong magnetic field. 
In order to further explore the consistency of the fall-back disk model with the X-ray spectra of AXPs and SGRs we embarked on a systematic study of their broad-band X-ray spectra, taking advantage of the current generation of hard X-ray telescopes (\suz and \int), and employing the best available spectral models  for the X-ray spectra from the accretion flow onto an accreting pulsar. In this paper we present the spectral analysis of the 0.5--200\,keV  spectra of 5 AXPs/SGRs with spectral models appropriate for modeling the accretion flow onto a pulsar.

The structure of this paper is as follows: in \S 2 we present the sample and data used in this study; in \S 3 we describe the analysis of these data, and the  results from the analysis of the X-ray spectra;  and in \S 4 we discuss the implications of our  results for the accretion model of pulsars. Finally in \S 5 we summarize our findings.   
All quoted errors correspond to the 90\% confidence limit for one interesting parameter. 

\section[]{The sample and  data}

 In order to obtain useful constraints on the X-ray spectral properties of AXPs/SGRs we selected to study objects from the list of known AXPs/SGRs  with available X-ray spectra above 10.0 keV (e.g. Mereghetti 2008; Enoto \etal 2010).  We searched the \suzaku\ (Mitsuda \etal 2007) and INTEGRAL (Winkler \etal 2003) archives for available data. 
 We found 5 objects with good quality spectra above 10\,keV. These are listed in Table \ref{T_sample}  along with the basic information of the observations used in this study. One additional object with known hard X-ray emission (SGR1806-20) was excluded from our analysis because it is near the Galactic Center and its hard X-ray emission might be significantly contaminated by the Galactic Ridge emission (Krivonos \etal 2007).

\begin{table*}
 \centering
\caption{Sample and data used in this study\label{T_sample}}
\begin{tabular}{@{}lccccccc@{}}
 \hline
Object & RA  & Dec & \multicolumn{2}{c}{\suzaku}             & INTEGRAL \\
       &     \multicolumn{2}{c}{(J2000)} & Obs Date  & Exposure  & Obs. Period  & Exposure \\
       &   hh mm ss.s  &  dd mm ss.s   &       & XIS, PIN  &         &        \\
       &     &     &       & (ksec)      &          &  (Msec)    \\
\hline
1E 1547-5408      & 15 50 54.1    &  -54 18 23.8  &    2009-01-28   &  10.6, 33.4         & \ldots             &  \ldots     \\ 
1E 1841--045      & 18 41 19.3    &  -04 56 11.2  &    2006-04-19   &  97.9, 63.4         & 2003-03 -- 2009-11 & 2.341       \\
4U 0142+61        & 01 46 22.4    &  +61 45 03.3  &    2009-08-12   &  107.4, 99.7	 & 2002-12 -- 2008-04 & 1.35       \\
1RXS J1708-4009        & 17 08 46.9    &  -40 08 52.4  &    2009-08-23   &  60.9, 47.9      & 2003-02 -- 2010-03 & 2.7         \\
SGR 1900+14       & 19 07 14.3    &  +09 19 20.1  &    2006-04-01   &  21.7, 14.3         & 2003-03 -- 2010-03 & 3.1         \\
\hline
\end{tabular}

 Positions  from Mereghetti (2008); the exposure times are corrected for deadtime.
\end{table*}

\subsection{\suzaku\ data}

 The \suzaku payload includes two X-ray detectors: the X-ray Imaging Spectrometer (XIS; Koyama \etal 2007),  and  the Hard X-ray Detector (HXD; Takahashi \etal 2007, Kokubun \etal 2007).
 The XIS consists of four identical CCD-based instruments (three of which are operational),  which provide imaging and spectroscopic data in the 0.5-10 keV band. It is placed at the focal point of the X-ray Telescope, which provides focused images with a half-power diameter of 2\arcmin\, (Koyama \etal 2007).
 The HXD  consists of two instruments: a PIN silicon-diode detector sensitive in the 10-60\,keV band, and  a GSO/BGO phoswich scintilator sensitive above 40\,keV. Being a non-imaging instrument, it uses collimators in order to reduce the background. 
 The lack of imaging information complicates the analysis of data for sources close to the Galactic Ridge since its contribution to the spectrum of the sources  can only be modeled. 

  The \suzaku data were obtained from the HEASARC archive.
  For their analysis  we followed the standard
procedures described in the \suzaku ABC guide\footnote{http://heasarc.gsfc.nasa.gov/docs/suzaku/analysis/abc/} v. 5.0, using  the FTOOLS
v.6.11 data analysis suite. The XIS data were first reprocessed and  screened using the
{\tt aepipeline} FTOOL with the default parameters. Then we extracted
lightcurves for each of the 3 available XIS detectors (XIS0, XIS2,
XI3) in order to search for strong background flares. None of the
sources showed any significant flaring that would require additional
screening.

Then we extracted images for each of the 3  XIS detectors based on which
 we defined source and background apertures. The source apertures were
selected to include as many as possible of the source photons, and in
general their radius was $\sim3$ arcminutes. Local background was
selected from circular regions adjacent to the source, while taking care to
avoid spillover of source counts due to the wings of the Point Spread
Function (PSF).  From those regions we extracted the source and
background spectra used in our analysis. Response matrices for the
source spectra were calculated using the {\tt xisrmfgen} and {\tt xissimarfgen} FTOOL.

The HXD data were reprocessed to apply the latest calibration files,
and screened using the {\tt aepipeline} FTOOL. The standard screening
criteria listed in the \suzaku ABC guide were applied. 
 Since the HXD is not an imaging instrument, an estimate of the X-ray
and particle backgrounds is based on detailed modeling of the in-orbit
background data (Fukazawa \etal 2009). 
 The final background spectrum was calculated from
the non-X-ray background data  produced by ISAS for
each observation sequence, and a model of the cosmic X-ray background (Moretti \etal 2009) using the {\tt hxdpinxbpi} FTOOL.

\subsection{\int data}

 The International Gamma-Ray Astrophysics Laboratory (INTEGRAL)  
carries  the  Imager on Board the INTEGRAL Satellite (IBIS; Ubertini et al. 2003) and the Spectrometer on INTEGRAL (SPI; Vedrenne et al. 2003). These instruments use coded aperture masks allowing the  reconstruction of X-ray images in the hard X-ray band.
The IBIS instrument in particular  consists of a CdTe detector array (ISGRI; Lebrun et al. 2003) and a CsI detector array (PICsIT). The ISGRI detector that will be used in this study,  is sensitive in the 20\,keV - 1\,MeV energy range, and provides a field of view of 19\deg$ \times$ 19\deg with a spatial resolution of $\sim12'$. The ISGRI detector is more sensitive than PICsIT, but overall it is less sensitive than the \suzaku PIN detector that operates in their overlapping energy range. 

We obtained the \int\ ISGRI spectra and response matrices
from the HEAVENS database (Walter \etal\ 2010), which provides data products  for known
sources from the analysis of all publically available INTEGRAL data up to the time of download (in our case September 2013).
 However, since these data are taken over the course of several years and generally 
 at a different epoch from the
\suzaku data, there is always the possibility for spectral or intensity
variations. For this reason we generally prefer to use the \suzaku XIS and HXD data. 
 When there is agreement between the spectra of \suz\ XIS and INTEGRAL ISGRI we use both in our analysis in order to improve the statistics of our spectra and to benefit from the extended coverage to higher energies provided by ISGRI.

\section{Spectral Fits}

\subsection{Description of the physical model}

 Since, according to the fall-back disk model, the energy output of AXPs/SGRs is generated by accretion of material onto the pulsar, one would expect X-ray spectra similar to those of X-ray pulsars. The expected accretion rates from a fall-back disk ($\sim10^{15}-10^{16}\rm{g \quad s^{-1}}$; Ertan \etal 2009)  are in the range of accretion rates of low-luminosity accreting pulsars. In this case, one would expect a power-law spectrum with a high-energy cutoff, which results from thermal and bulk-motion Comptonization in the accretion flow (e.g.  Lyubarskii \& Sunyaev 1982; Becker \& Wolff 2007, hereafter BW07; Kylafis \etal 2014). The slope of the power law and the energy of the cutoff depend on the energy of the seed photons, the temperature of the electrons, the velocity of the accretion flow, and the optical depth in the accretion column, which determines the number of scatterings that a photon will undergo before emerging (e.g. BW07; Kylafis \etal 2014).  

 The spectral model required to fit the data corresponds to the following physical picture (first discussed in Tr\"{u}mper \etal 2010, 2013; Kylafis \etal 2014): photons produced in the accretion flow are up-scattered by the thermal as well as  the bulk-motion Comptonization mechanisms by the infalling electrons. These up-scattered photons comprise the power-law component of the observed spectrum. 
In addition to the non-thermal emission from the accretion flow,  thermal photons from the polar cap of the neutron star also contribute to the observed spectrum, either directly, or after beeing comptonized by electrons at the atmosphere of the polar cap (c.f. Ferrigno et al. 2009).

   Modeling  the X-ray spectra produced by accretion of matter onto a pulsar is a very complex and still open problem. The best currently available models for the spectra produced in the  accretion flow are based on the treatment of Becker \& Wolff  (Becker \& Wolff 2005; BW07) which  was implemented in XSPEC by Ferrigno \etal (2009). This model used the Green's function formalism to solve analytically the radiative transfer equation within a cylindrical accretion flow in the presence of a strong magnetic field. 

 As discussed in BW07, the photons produced at or below the accretion shock are trapped in the accretion flow and they can only escape sideways. The optical depth in the transverse direction is of the order of a few (e.g. BW07; Tr\"{u}mper \etal 2013; Kylafis \etal 2014), therefore the photons escaping from the sides of the accretion column have undergone several scatterings and they have increased their energy. 
This results in a hard power-law spectral component  associated with the fan beam. 
On the other hand, the cross-section along the magnetic field axis is  (BW07)

\[
 \sigma(E) = \left\{ 
  \begin{array}{l l}
     \sigma_{T}(E/E_{C})^{2} & {\rm{if}} \quad E<E_{C} \\
     \sigma_{T}			    &  {\rm{if}} \quad E>E_{C} \\
\end{array}  \right.
\]
where   $E_{C}=11.6B_{12}$\,keV is the cyclotron energy, and  $B_{12}$ is the magnetic field at the surface of the neutron star in units of $10^{12}$\,G.
Since the magnetic field along the magnetic axis decreases (as $B\propto z^{-3}$), the energy of the resonant cyclotron  scattering cross-section will decrease accordingly. This makes the vertical direction of the accretion flow (i.e. parallel to the dipole magnetic field axis) optically thick to all photons produced at the thermal mound or  below the accretion shock.

  The Ferrigno \etal (2009) model includes approximations for the angle-averaged scattering  cross-sections in the parallel and perpendicular directions with respect to the magnetic field, and it accounts also for the effects of thermal and  bulk-motion Comptonization (e.g. Titarchuk \etal 1997), the latter being particularly important in this accretion flow geometry. It accounts for the three main emission mechanisms that operate within the accretion flow: thermal (black-body) emission at the base of the accretion column (thermal mound), cyclotron emission due to  transitions from the first Landau level of the  collisionally excited  electrons in the strong magnetic field of the pulsar (also including the inverse process of resonant cyclotron absorption), and  bremsstrahlung radiation from the electrons. However, it has several shortcomings such as decoupled treatment of the thermal and bulk Comptonization, various approximations for the relation between parameters of the radiative transfer  and the geometry of the accretion flow (e.g. parameterization of the velocity of the accretion flow in terms of the optical depth), angle-averaged scattering magnetic cross sections, and a simple cylindrical accretion flow. Despite these shortcomings, until recently,  this was  the best available model  for analyzing the spectra of accreting pulsars.

More recently,  Farinelli \etal (2012; hereafter F12), following the basic formulation of the radiative transfer equation by  BW07,  proposed another method to calculate the spectra from the accretion column,   that overcomes some of the limitations of the Ferrigno \etal (2009) implementation. More particularly, this new solution  allows for a more flexible treatment of the velocity profile of the accretion flow by allowing the velocity to be a function of the height rather than a function of the optical depth, and most importantly reduces the number of fitted model parameters  that are strongly correlated.

 Therefore, in our parameterization of the spectrum, we combine  the F12 model which describes the emission from the accretion column, and a black-body component which describes the   direct  thermal emission from the polar cap. One would expect that a fraction of the photons emitted by the polar cap would interact with the accretion flow and be Comptonized, contributing to the hard X-ray emission.  However, as discussed in Kylafis \etal (2014),  for a cylindrical geometry of the accretion flow, the  fraction of these itersected photons is very small and they  hardly contribute in the hard X-ray spectra, supporting the decoupled modeling of the thermal emission from the polar cap  and the emission from the accretion column that is adopted in this study. 

 The F12 model is parameterized in terms of several physical parameters: the temperature of black-body photons at the thermal mound at the base of the accretion flow ($kT\rm{_{bb}^{TM}}$), the velocity profile of the accretion flow, the temperature of the electrons in the accretion flow responsible for the thermal Comptonization ($kT_{e}$), the total vertical optical depth along the accretion column (i.e. parallel to the magnetic field lines; $\tau$), and the radius of the  accretion column ($r_{0})$. 

 For the velocity profile, the F12 model allows for two possibilities.  
One is the velocity profile adopted by BW07, which is
 
    \begin{equation}  
    \beta(\tau) = -\Psi \tau,
    \end{equation} 
where $\beta$ is the velocity of the infalling material in terms of the speed of light, $\Psi$ is a proportionality factor, and $\tau$ is the vertical optical depth.
 This means that the magnitude of the velocity is maximum at the top 
of the accretion column, i.e. at the place which BW07 call ``sonic surface'' 
(see their Fig. 1), and zero at the surface of the neutron star.  Such a 
velocity profile is appropriate for high-luminosity X-ray pulsars, in which 
the radiative shock occurs far from the neutron star surface. In this model,
the flow {{\it above}} the ``sonic surface'' is not taken into account.

 The second velocity profile allowed in the F12 model is the
free-fall one:

    \begin{equation}\label{Eq:vel}  
    \beta(z) = -A (\frac{z_{s}}{z})^{\eta}, 
    \end{equation}
 where $z \ge z_{s}$ is the distance from the center of the neutron star,
$z_s$ is the distance of the sonic point in the accretion flow from the 
center of the neutron star (c.f. Basko \& Sunyaev 1976), 
$\eta$ is the index of the velocity profile ($\eta=0.5$ for free fall), 
and  $A$ is a normalizing constant defined as 
 $A=\beta_{0}(R/{z_{s}})^{\eta}$, where $\beta_{0}$ is the magnitude 
that the velocity, in units of the speed of light, would aquire at the 
surface of the neutron star of radius $R$.
Such a velocity profile is appropriate for low-luminosity X-ray pulsars,
in which the radiative shock occurs near the neutron star surface.  In this
model, the flow {{\it below}} the sonic point is not taken into account.

 In our analysis, we use the second velocity profile, because it is 
more appropriate for the low luminosities involved in AXPs and SGRs.

 Another parameter of the F12 model is the albedo of the surface at the base of the accretion flow, which for a neutron star is fixed to 1 (a fully reflective surface). 
 The overall normalization of the model is parameterized in terms of the luminosity of the seed black-body emitting surface at the thermal mound. In addition, by energy conservation the observed luminosity is equal to the gravitational energy released in the accretion flow and proportional  to the mass accretion rate. 

 In order to account for emission from the polar cap outside the accretion flow, in addition to the F12 accretion flow spectral model ({\tt compmag}) we  include in our model  one more  black-body component ({\tt bb}). The temperature of the black-body component describing the emission from the polar cap ($kT\rm{_{bb}^{PC}}$) was initially tied to the temperature of the black-body seed photons from the thermal mound in the {\tt{compmag}} model ($kT\rm{_{bb}^{TM}}$). However, we also tested if allowing the two black-body components to have different temperatures significantly improves the fit. This was the case for the spectra of 4U 0142+61 and RXJS\,1708-4009, and the two parameters were fitted independently. In the other 3 sources the two black-body temperatures were tied together. 
 Finally, we included photoelectric absorption ({\tt phabs} model in XSPEC) in order to account for absorption by interstellar material along the line of sight.  Therefore, our adopted model consists of two additive components seen through the same absorber: {\tt{phabs * (bb + compmag)}}.

\subsection{Spectral analysis}

  The spectral analysis was performed with the XSPEC v12.8  spectral-fitting package (Arnaud \etal 1996), using the implementation of the F12 model  (model {\tt compmag}).  The spectral data were binned  to have at least 20 net counts in each bin in order to allow the use of $\chi^{2}$ statistics.  
   Before analyzing the spectra, we fitted them with the best-fit
model published by Enoto \etal (2010) in order to check the
consistency of our processing of the \suzaku\ data. In all cases we found consistent broad-band fluxes and
spectral slopes above 10 keV  with those reported in Enoto \etal (2010).

In the case of 1E 1841--045, which is located within a bright Supernova Remnant (SNR), we included in the spectral model a thermal plasma component (APEC; Smith \etal 2001) in order to account for the thermal emission from the SNR. In addition, the spectrum of 4U 0142+61 showed a strong artifact at an energy of 1.84\,keV (possibly due to the Si K$\alpha$ absorption edge) that could not be modeled adequately. For this reason we excluded the 1.7-2.0\,keV energy range from the spectral fit. 

\begin{figure*}
\includegraphics[angle=270,width=8cm]{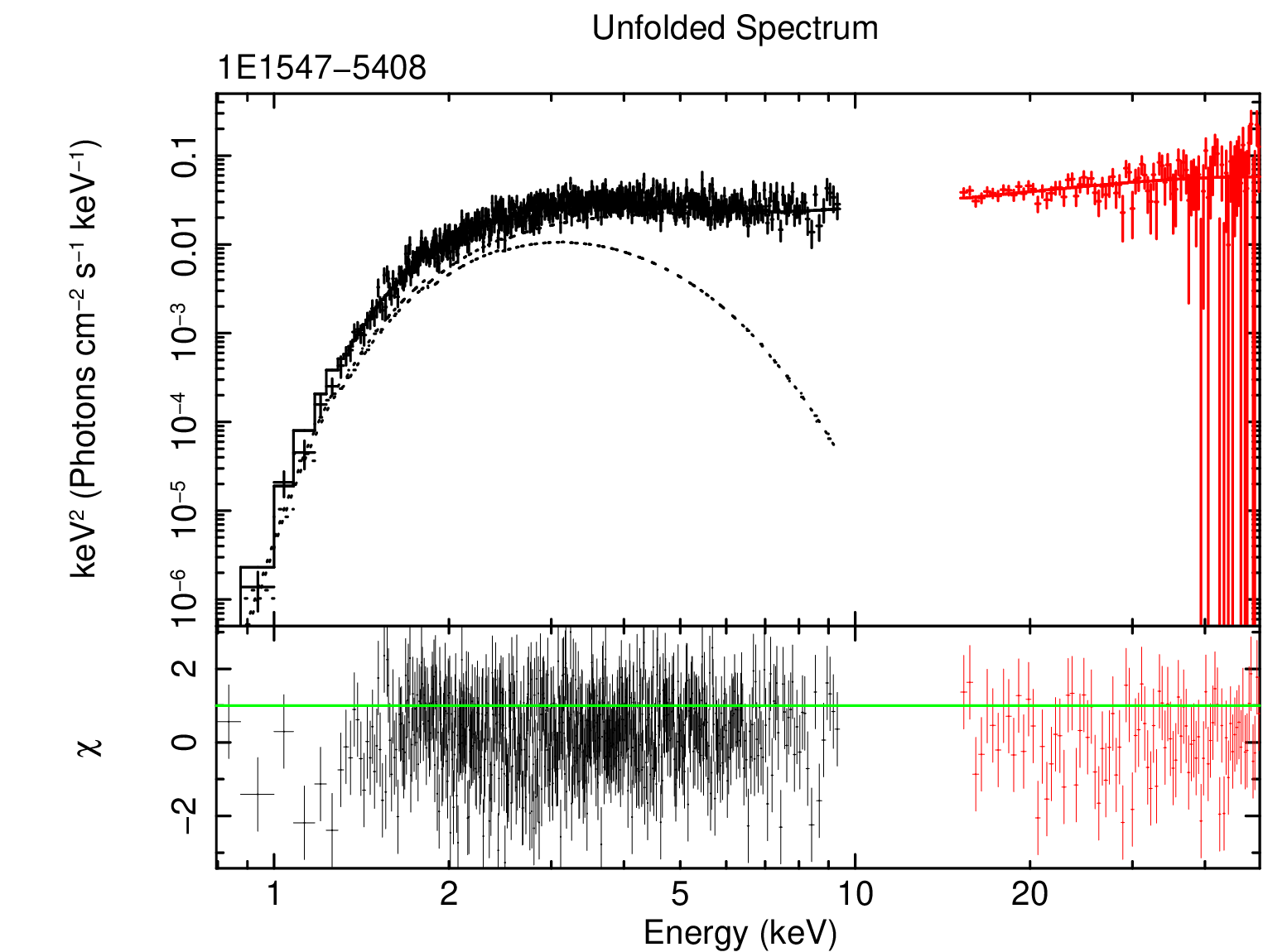}
\includegraphics[angle=270,width=8cm]{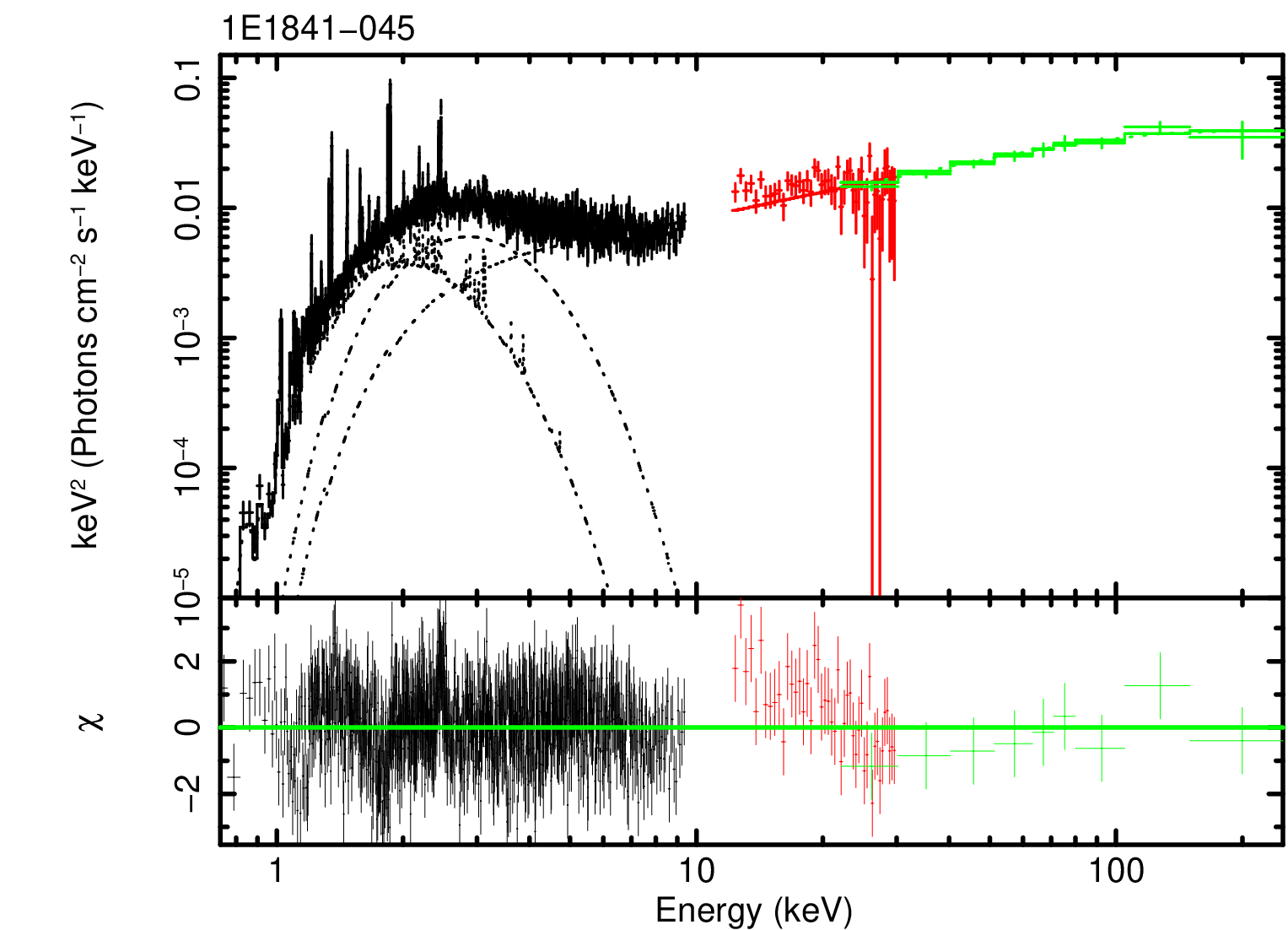} 
\includegraphics[angle=270,width=8cm]{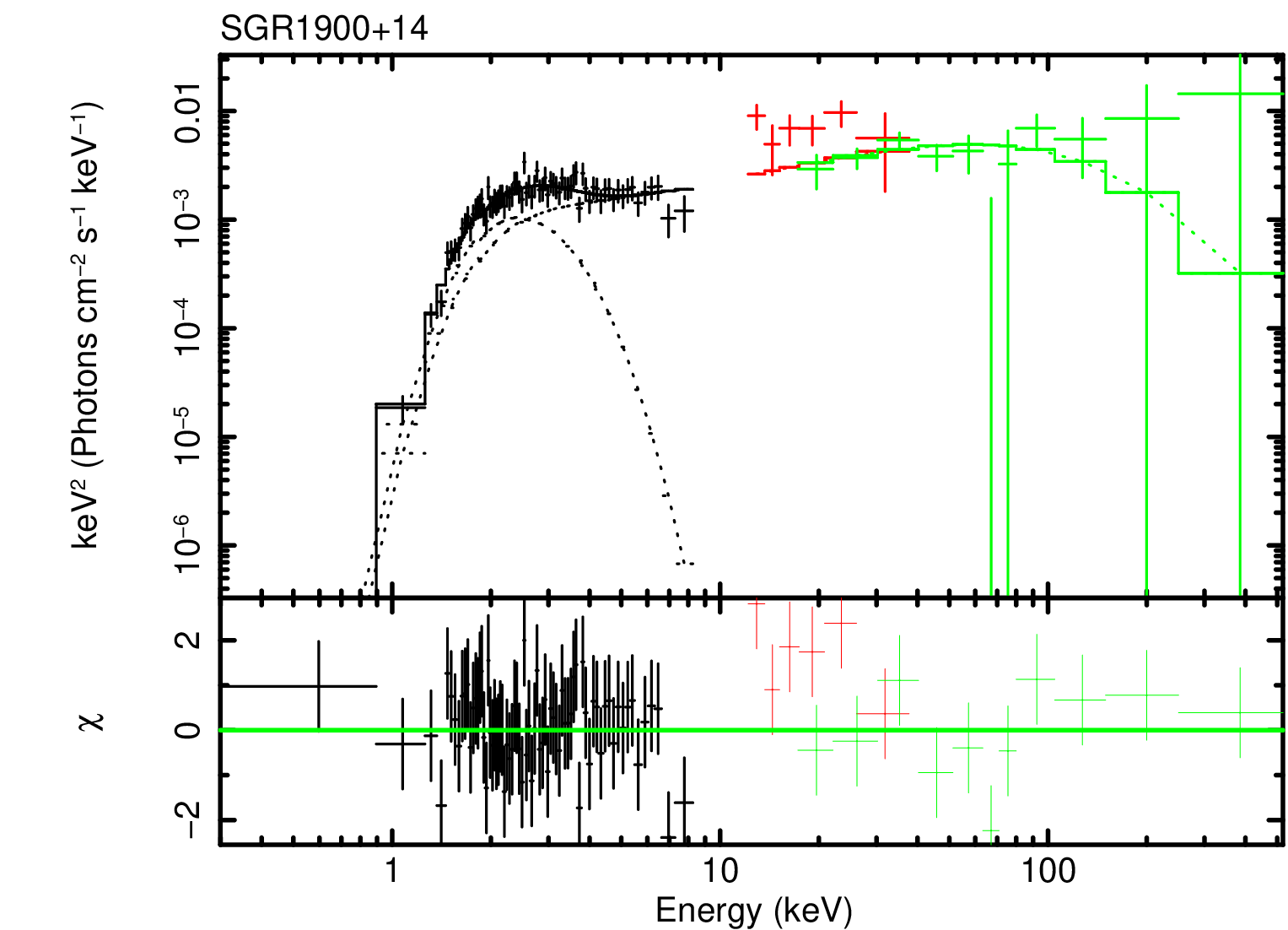}
\caption{The \suzaku (XIS and HXD) and INTEGRAL (ISGRI) spectra of 1E 1547-5408, 1E 1841--045, and   SGR1900+14,
 (in $\rm{E^{2}\frac{dN}{dE}}$ scale)  along with their best-fit models presented in
Table \ref{T_sp_fits}. The \suzaku XIS and HXD spectra are shown in black and red respectively, while the \int ISGRI spectra are shown in green. 
For clarity we only show spectra from one XIS detector.
 In the case of 1E 1547-5408 the quality of the
INTEGRAL  ISGRI spectra was poor and only the \suzaku HXD-PIN spectra
are shown. The contribution of the  black-body component representing the polar cap emission, and the {\tt compmag} accretion flow component to the overall spectral model (solid line),  are shown by the dashed lines (the spectrum of 1E 1841-045 also includes a thermal plasma component  shown by the dash-dot line). The bottom
panel of each figure shows the residuals from the best-fit model (in $\sigma$).\label{F_sp_fits}}
\end{figure*}

 The absorbed {\tt{compmag + bb}} model gave good fits to the data (reduced $\chi^{2}<1.2$). The best fit parameters along with their 90\% confidence intervals for one interesting parameter ($\Delta\chi^{2}=2.709$) are presented in Table \ref{T_sp_fits}. 
 The unfolded spectra along with the best-fit models  in $E^{2}(dN/dE)$ space are presented in Fig. \ref{F_sp_fits}. In the same figure we show the fit residuals in terms of the error of each data point (in $\sigma$). 
These fits were performed with the parameter $\beta_{0}$ (corresponding to the  magnitude of the velocity of the accretion flow, in units of the speed of  light, at the neutron star surface; Eq. \ref{Eq:vel}) fixed to $\beta_{0}=0.6$, as expected for free-falling material onto 
a neutron star of mass 1.4\msun\  and radius 12.5\,km.
 We also performed spectral fits with $\beta_{0}$ free to vary. In all but one case (1E 1547-5408) this parameter pegged at the maximum value ($\beta_{0}\simeq 1.0$), while the best-fit  electron temperature $kT_{e}$ was unphysically low ($\sim1$\,keV). This, in combination with the marginal improvement of the fit quality over the fits with fixed $\beta_{0}$,  led us to adopt the  best fit 
results with  $\beta_{0}$ fixed to 0.6, for all sources apart from  1E 1547-5408 (however, fixing $\beta_{0}$ to 0.6 in 1E 1547-5408 also gives an acceptable fit).

\begin{figure*}
\includegraphics[angle=270,width=8cm]{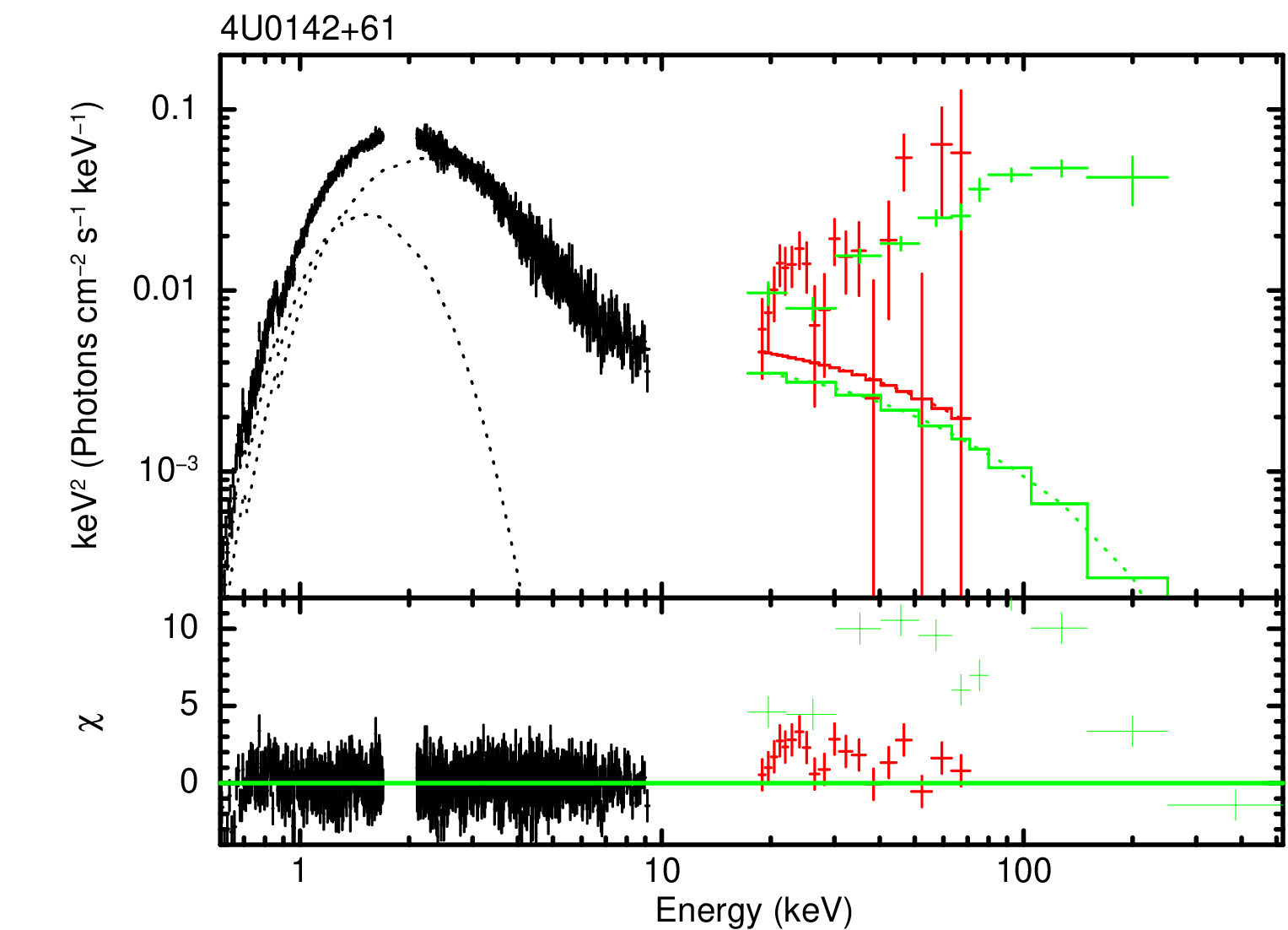}
\includegraphics[angle=270,width=8cm]{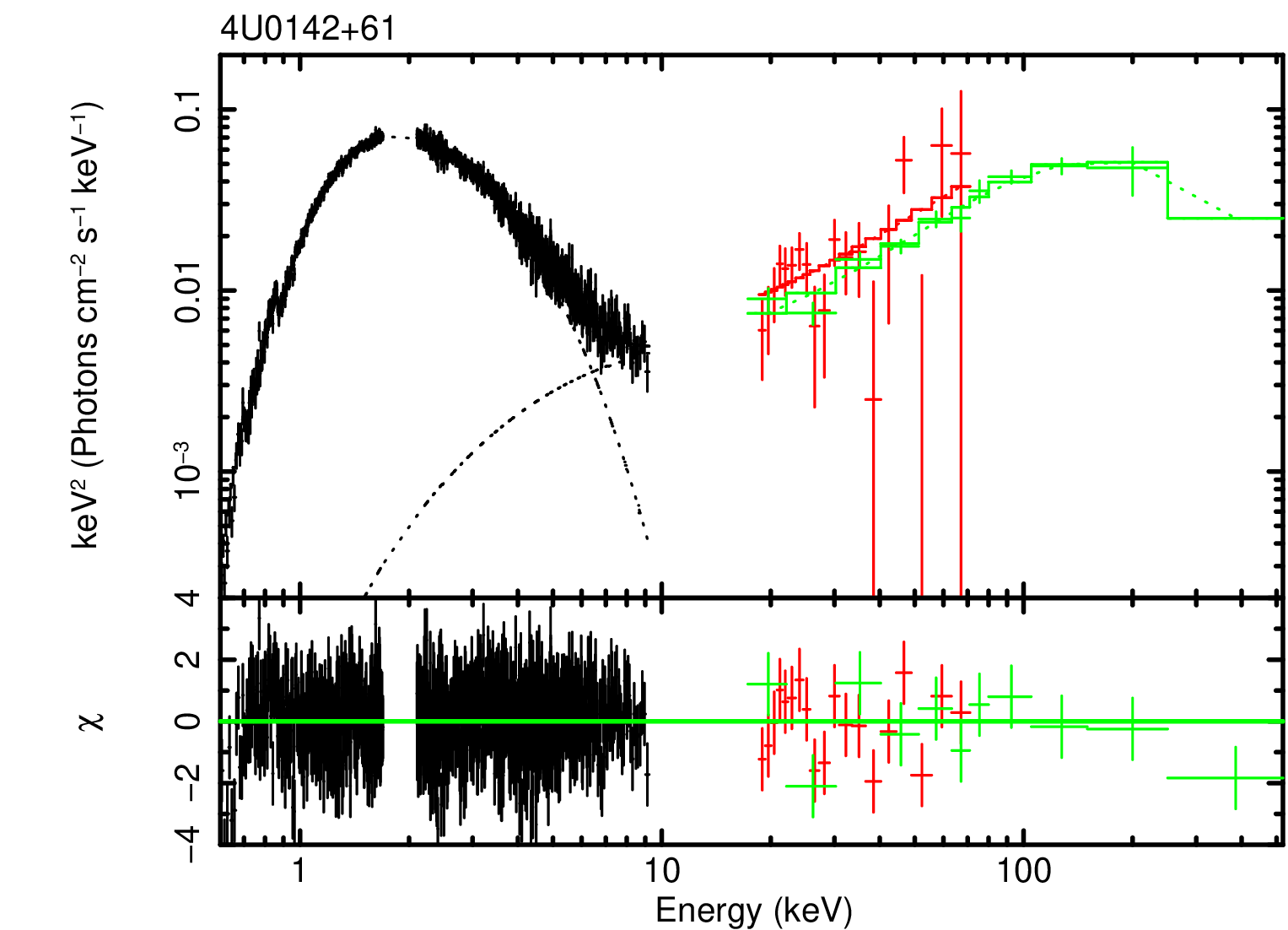}
\includegraphics[angle=270,width=8cm]{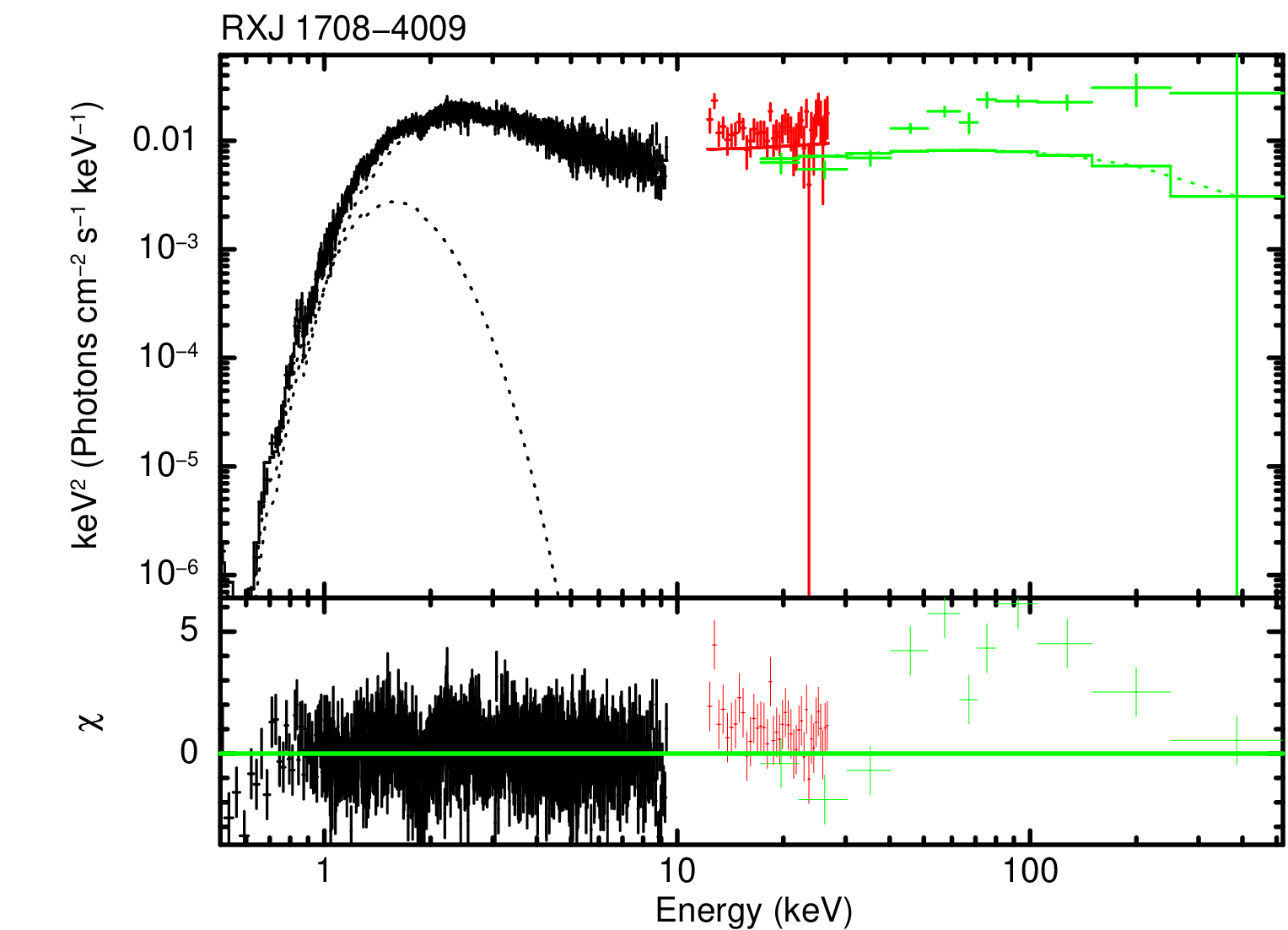}
\includegraphics[angle=270,width=8cm]{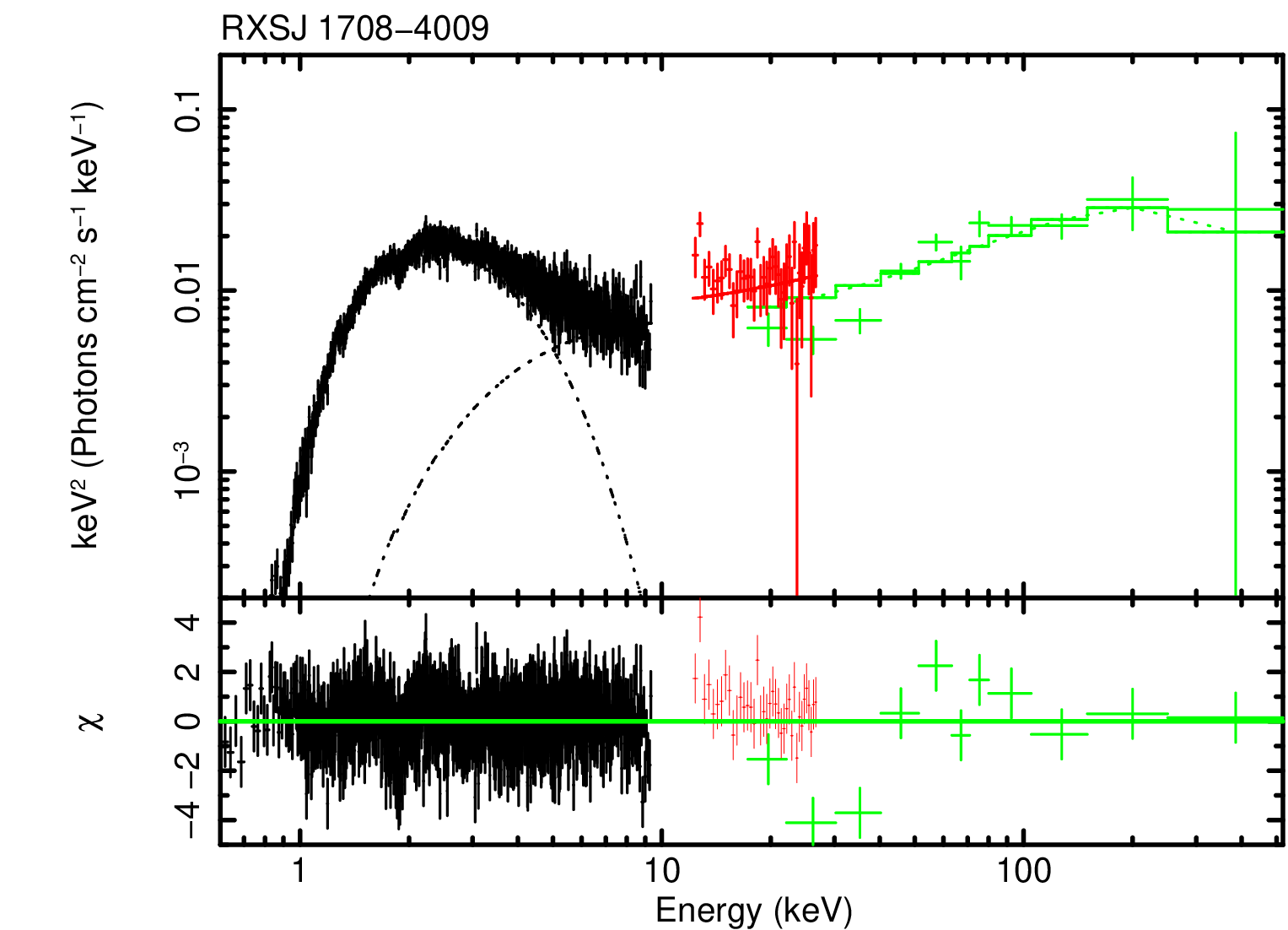}
\caption{The \suzaku (XIS and HXD) and INTEGRAL (ISGRI) spectra of 4U0142+61 and 1RXS J1708-4009 (in $\rm{E^{2}\frac{dN}{dE}}$ scale) fitted with the {\tt{phabs*(bb + compmag)}} (left; Table \ref{T_sp_fits}) and the  {\tt{phabs*(comptb + compmag)}} (right; Table \ref{T:4U0142_fit}) models.  
The \suzaku XIS and HXD spectra are shown in black and red respectively, while the \int ISGRI spectra are shown in green.  
For clarity we only show spectra from one XIS detector.
The spectra have been corrected for instrumental
effects. We clearly see how the addition of the thermal Comptonization model improves the fit. The bottom
panel of each figure shows the residuals from the best-fit model (in $\sigma$).\label{F:4U0142_fit} }
\end{figure*}

In the cases of 4U 0142+61   and 1RXS J1708-4009 (see Fig. \ref{F:4U0142_fit}, left), we see that the adopted model does not adequately reproduce the hard tail of the spectrum. In order to remedy this, we included in our model a thermal Comptonization component. Such a component could describe the Comptonization of the thermal emission from the neutron-star surface by an atmosphere of hot electrons (e.g. Ferrigno \etal. 2009). We  used the model {\tt{comptb}} in XSPEC (Farinelli \etal 2008; hereafter F08), with the parameter $\delta$ fixed to 0 (which corresponds to a pure thermal Comptonization spectrum). This model is parameterized in terms of the temperature of the seed black-body photons ($kT\rm{_{bb}^{TC}}$), which we allowed to differ from the temperature of seed black-body photons in the accretion flow,  a power-law slope modifying the seed black-body spectrum (which we fixed to $\gamma=3$, i.e. a pure black body), the index of Green's function $\alpha$, the temperature of the Comptonizing electrons ($kT\rm{_{e}^{TC}}$), and the relative contribution of the seed black-body photons and upscattered photons to the total spectrum (the illuminating factor, $\log(A)$). 
A spectral fit with a combination of the {\tt{comptb}}, the  {\tt{compmag}}, 
 and the {{\tt bb}} models (model {{\tt{phabs*(comptb+compmag+bb)}}} 
in XSPEC) greatly improved the fit, particularly at the hard part of the spectrum, while it gave more physical values to the other model parameters. In order to limit the number of free parameters we fixed the terminal velocity of the accretion flow to 0.6c, as with the spectral fits for the other sources. The model fit along with the residuals are shown in Fig. \ref{F:4U0142_fit} (right), while the best fit parameters are presented in Table \ref{T:4U0142_fit}.  Hereafter we consider this as the best fit model for 4U 0142+61  and 1RXS J1708-4009, and all derived parameters will be based on its best-fit parameters.   

\begin{table*}
\centering
\caption{Spectral fit results\label{T_sp_fits}}
\begin{tabular}{@{}lccccc@{}}
\hline
Parameters & \multicolumn{4}{c}{Objects} \\
                                              &  1E 1547-5408       & 1E 1841--045$^{\dag}$  & 4U 0142+61& 1RXS J1708-4009 & SGR1900+14\\
\hline
\nh    \,   $(10^{22}\rm{cm^{-2}})$           &$2.91^{+0.12}_{-0.08}$     & $2.66\pm0.1$              &  $0.68\pm0.01$         &    $1.61\pm0.06$   & $2.6\pm0.4$   \\[3pt]
$kT\rm{_{bb}^{PC}}$ (keV)                     &$0.61\pm0.03$              & $0.59\pm0.01$             &  $0.275\pm0.004$       &    $0.197\pm0.005$ & $0.41^{+0.12}_{-0.09}$   \\[3pt]
Norm$\rm{_{bb}}$ $(\times10^{-4}L_{39}/D^{2})$&$4.76\pm0.50$              & $3.01\pm0.02$             &  $15.9^{+0.2}_{-0.5}$  &    $9.9\pm0.3$     & $0.76^{+0.32}_{-0.29}$   \\ [3pt]
$kT\rm{_{bb}^{TM}}$ (keV)                     &$0.61^{*}$                 & $0.59^{*}$                & $0.436^{+0.019}_{-0.005}$ & $0.397\pm0.006$ & $0.41^{*}$   \\[3pt]
$kT\rm{_{e}}$$^{\P}$ (keV)                    &$0.23_{-0.2}^{+3.6}$       & $60_{-22}^{+20}$          & $0.012^{+0.007}_{-0.001}$&  $0.02\pm0.01$   & $0.2 (<31.7)$  \\[3pt]
$\tau_{T}$$^{\ddag}$                          &    $170^{+60}_{-10}$      & $156^{+48}_{-17}$         &  $104^{+17}_{-2}$      &    $160\pm3$       & $263^{+75}_{-60}$   \\[3pt]
$\beta\rm{_{0}}$ (c)                          &$0.47_{-0.08}^{+0.02}$     & $0.6$ (f)                 &  $0.6$  (f)            &    $0.6$ (f)        & $0.6$   (f) \\[3pt]
$r\rm{_{0}}$ (m)                              &$272^{+4}_{-8}$            & $2800 (>250)$             &  $169\pm2$             &    $183.4\pm0.4$   & $263^{+76}_{-60}$            \\[3pt]
Norm$_{compmag}$$^{**}$                        &$321_{-192}^{+68}$         & $96_{-23}^{+44}$         &  $17600^{+2600}_{-5100}$ &    $7227\pm51$    & $131^{+288}_{-42}$  \\[3pt]
$\chi^{2}/dof$                                &1398.7/1341                & 1112.9/962                &  1034/919              &    1251/1078       &  85.2/85    \\[3pt]
\hline
\end{tabular}

$\dag$  The model for the spectral fit of 1E 1841-045 includes an APEC thermal plasma component (e.g. Smith \etal 2001; Foster \etal 2012). Its best-fit  temperature and abundance are $kT=0.51\pm0.02$\,keV, and $Z=0.40\pm0.07$ respectively (c.f. Fig.\ref{F_sp_fits}).\\
$^{*}$  The temperature of the black-body component is fixed to the temperature of the black-body parameter of the {\tt{compmag}} component. \\ 
$\P$ The temperature of the electrons in the accretion flow; see \S4.2 for a detailed discussion of the fit results.\\
$\ddag$  The Thompson optical depth along the accretion flow (parallel to the magnetic field) (with a conversion factor of  $\sigma_{||}=10^{-3}\sigma_{T}$). \\ 
$**$   The normalization of the {\tt compmag} model is defined as $R^{2}_{\rm{km}}/D^{2}$, where $R_{\rm{km}}$ is the radius of the black-body emitting area, and $D$ is the distance of the source in units of 10\,kpc. \\ 
(f) Fixed parameter. 
\end{table*}

\begin{table*}
\centering
\caption{Spectral fit results for 4U 0142+61 and 1RXS J1708-4009\label{T:4U0142_fit} }
\begin{tabular}{@{}lcc@{}}
\hline
Parameter                                      & \multicolumn{2}{c}{Fit results} \\
                                               & 4U 0142+61                 & 1RXS J1708-4009   \\
\hline
\nh    \,   $(10^{22}\rm{cm^{-2}})$            &  $0.67\pm0.01$            &  $1.76^{+0.19}_{-0.13}$   \\[3pt]
$kT\rm{_{bb}^{TC}}$ (keV)                      &  $0.28_{-0.01}^{+0.04}$   &  $0.16_{+0.02}^{-0.03}$   \\[3pt]
Norm$\rm{_{TC}}$ $(\times10^{4}L_{39}/D^{2})$ &  $29.6^{+0.7}_{-0.5}$     &  $22.7^{+0.2}_{-0.1}$     \\[3pt]
$\gamma\rm{_{bb}^{TC}}^{\P\P}$                 &  3.0                      &  3.0                      \\[3pt]
$\alpha\rm{^{TC}}$                             &  $2.27^{+0.14}_{-0.69}$   &  $1.4^{+0.3}_{-0.2}$      \\[3pt]
$\delta\rm{^{TC}}^{\P\P}$                      &  0.0                      &  0.0                      \\[3pt]
$\log(A)\rm{^{TC}}$                             &  $4.6^{+0.4}_{-4.4}$      &  $0.06^{+0.54}_{-0.35}$   \\[3pt]
$kT\rm{_{e}^{TC}}$$^{\dag}$ (keV)              &  $0.88^{+0.15}_{-0.05}$   &  $0.74^{+0.11}_{-0.08}$   \\[3pt]
$\tau\rm{_{T}}$$^{\ddag}$            &  $53^{+356}_{-23}$        &  $39^{+35}_{-27}$         \\[3pt]
$\beta\rm{_{0}}$ (c)$^{\P\P}$                       &  $0.6$                    &  $0.6$                    \\[3pt]
$kT\rm{_{bb}^{TM}}$ (keV)                      &  $1.42_{-0.09}^{+0.18}$   &  $1.2_{+0.6}^{-0.3}$      \\[3pt]
$kT\rm{_{e}}$$^{\P}$ (keV)                     &  $22^{+2}_{-5}$           &  $40.2^{+50}_{-16}$       \\[3pt]
$r\rm{_{0}}$ (m)                                    &  $329\pm9$                &  $231^{+412}_{-82}$       \\[3pt]
Norm$_{compmag}$$^{*}$                         &  $1.7^{+18.8}_{-0.13}$    &  $7.2^{+22.6}_{-4.4}$     \\[3pt]
$\chi^{2}/dof$                                 &  1001/919                 &  1258/1106                \\
\hline
\end{tabular}

$\P\P$ Model parameter fixed. \\
$\dag$  The temperature of the electrons in the {\tt comptb} model.\\
$\ddag$  The Thompson optical depth along the accretion flow (parallel to the magnetic field)  (with a conversion factor of  $\sigma_{||}=10^{-3}\sigma_{T}$). \\ 
$\P$ The temperature of the electrons in the accretion flow; see \S4.2 for a detailed discussion of the fit results.\\
$*$   The normalization of the {\tt compmag} model is defined as $R^{2}_{\rm{km}}/D^{2}$, where $R_{\rm{km}}$ is the radius of the black-body emitting area, and $D$ is the distance of the source in units of 10\,kpc. \\ 
\end{table*}

\section{Discussion} 

In the previous sections we have presented the results from the analysis of
the X-ray spectra of 5 AXPs and SGRs  with available data 
above 10\,keV.  
The main driver of this investigation was to test whether the  X-ray spectra of 5 AXPs/SGRs are consistent with the best available model for the X-ray spectrum expected from the accretion column onto a pulsar ($B\sim10^{12}$G; Farinelli \etal 2012). This model is better suited for the X-ray spectra of AXPs/SGRs than the models used in similar investigations in the past (e.g. Tr\"{u}mper \etal 2010; Guo \etal 2015), because of its more appropriate geometry and the use of (angle-averaged) magnetic cross-sections.  Furthermore, it has a physical basis that enables us to derive and assess the validity of the physical parameters of the accretion flow within the context of the fall-back  disk model, in contrast to phenomenological fits using power-law models.  

Our analysis shows that   the spectra of the 5 AXPs/SGRs studied in this work are well fitted with a combination of a black-body model and the accretion column model of F12 (only in the cases of 4U 0142+61 
 and 1RXS J1708-4009 we required an additional thermal Comptonization model).  Even though this model includes several assumptions and simplifications in its treatment of the accretion flow and radiative transfer (e.g. cylindrical geometry, angle averaged scattering magnetic cross sections, independent treatment of the thermal and bulk Comptonization)  it is currently the best available model to study the X-ray emission from accreting pulsars and allows us to: (a) perform a basic test of the accretion model, and (b) obtain a first picture of the physical parameters of the accretion flow.

 \begin{table*}
 \begin{centering}
\caption{Source Luminosity \label{T_Lum}}
\begin{tabular}{@{}lccccccc@{}}
\hline
Object & Distance$^{\dag}$ &\multicolumn{2}{c}{Flux}  & \multicolumn{2}{c}{Luminosity} \\
       &        &(0.1-10.0) keV  & (10.0-250.0) keV    & (0.1-10.0) keV   & (10.0-250.0) keV  \\
       &        &obs. (unabs.)  & obs.$^{\ddag}$    & obs.  (unabs.)   & obs.\\ 
       & kpc    &($10^{-11}$\funit)  & ($10^{-11}$\funit)    & ($10^{35}$\ergs)   & ($10^{35}$\ergs)  \\
\hline
1E 1547-5408         & 4.5   &  6.67 (12.66) &   19.2  &  1.6 (3.1)   &  4.6   \\   
1E 1841--04$^{*}$    & 8.5   &  2.13 (4.43)  &   8.52  &  1.8 (3.8)   & 11.0   \\  
4U 0142+61           & 3.6   &  13.8 (31.7)  &   1.3   &  2.1 (4.9) & 0.2    \\  
1RXS J1708-4009      & 3.8   &  4.12 (26.3)  &    8.2  &  0.7 (4.5)   &  1.4   \\  
SGR1900+14          & 12.5  &  0.54 (1.4)   &   1.86  &  1.0 (2.6)  &  3.5  \\  
\hline
\end{tabular}
\end{centering}

$\dag$ The adopted distances are from Olausen \& Kaspi, 2014. \\
$\ddag$ The unabsorbed hard X-ray flux is equal to the absorbed flux for the $\rm{H_{I}}$ column densities considered here.\\
$*$  The reported fluxes for 1E 1841--04 exclude the thermal emission from the surrounding Supernova Remnant. 
\end{table*}

 \begin{table*}
 \begin{centering}
\caption{Physical parameters of the accretion flow \label{T_phys_param}}
\begin{tabular}{@{}lccccccc@{}}
\hline
Object              & Dist.$\dag$ & $\dot{M}$ &$\tau_{\perp}$  & $\rm{R_{pc}}$  \\
                    & kpc         & $10^{15}\,\rm{gr s^{-1}}$     &              & km         \\
\hline
1E 1547-5408         & 4.5   &  5.2   &   1.4   &  0.77       \\
1E 1841--04          & 8.5   &  10.0  &   0.25   &  3.6       \\
4U 0142+61           & 3.6   &  3.4  &   0.7   &  18.3$^{\P}$       \\
1RXS J1708-4009      & 3.8   &  3.9   &   1.2   &  62.3      \\
SGR1900+14          & 12.5  &  4.4   &   1.2  &  5.4      \\
\hline
\end{tabular}
\end{centering}

$\dag$ The adopted distances are from Olausen \& Kaspi, 2014. \\
$^{\P}$ The radius of the polar cap for 4U 0142+61 and 1RXS J1708-4009 is based on the best-fit normalization and black-body temperature of the {\tt{comptb}} model and it is likely overestimated since this model may also account for some emission produced in the accretion flow.
\end{table*}

In the cases of 4U 0142+61 and 1RXS J1708-4009, the spectral fit requires 
an additional thermal Comptonization model, which we attribute to either 
the accretion flow  below the shock or Comptonization of  photons 
from the polar cap by electrons at its atmosphere (c.f. Ferrigno \etal 2009). 
The low temperature of the seed photons ($kT\rm{_{bb}^{TC}}\sim0.7-0.9$\,keV) and Comptonizing electrons ($kT\rm{_{e}^{TC}}\sim0.1-0.3$\,keV) suggests that the Comptonization does not take place in the accretion flow (in contrast the electron temperature in the accretion flow is $\sim22$\,keV). However, we cannot rule out a contribution from the accretion flow. 

  The analysis presented in \S3 shows that this model can describe very well the broad-band (0.5-200 keV) X-ray spectra of AXPs and SGRs, indicating that they could be produced from an accretion flow onto the pulsar.  
The spectral fits show that the temperature of the thermal emission from the polar cap  is
$kT\rm{_{bb}^{PC}}\sim0.2-0.6$\,keV. This is a factor of $\sim2$ higher than the typical temperatures of ``normal'' pulsars, but it is consistent with the polar cap temperature of other AXPs (e.g. Aguilera \etal 2008), and could be understood in terms of heating of the polar cap by the fan beam (Tr{\"u}mper \etal 2013).

\subsection{Physical parameters of the accretion flow}

 From the best-fit model  parameters we can derive some basic information about the physical conditions in the accretion column.  
 Since the energy released is gravitational, we have that the mass 
accretion rate is given by

     \begin{equation}  
     \dot{M} = \frac{L R}{GM}
    \end{equation} 
 where $L$ is the luminosity of the emitted radiation, $R$ and $M$ are the radius and the mass of the neutron star respectively, and $G$ is the gravitational constant. 
The mass accretion rates estimated from the luminosities listed in Table \ref{T_Lum} and assuming a neutron-star mass $M=1.4\,\rm{M_{\odot}}$ and radius $R=\rm{12.5\,km}$ are presented in Table \ref{T_phys_param}.  They are all in the range $\sim10^{15}\rm{g ~ s^{-1}}$, which is $\sim100-1000$ times lower than the accretion rates of typical accreting pulsars. They are also in good agreement with the fall-back disk  model which predicts accretion rates of $10^{15}-10^{16}\rm{g ~ s^{-1}}$ for $\sim10^{5}$yr after the supernova explosion (e.g. Ertan \etal 2009).

 The radius of the accretion column resulting form the spectral fits is generally between 160-330\,m.  Assuming that the accreted material is captured between the Alfv\'{e}n radius and the co-rotation radius, Kylafis \etal (2014) calculated that the radius of the accretion column is $r_{o}\approx 170\dot{M}^{1/5}_{15}B_{12}^{-1/4}\,\rm{m}$, where $\dot{M}\rm{_{15}}$ is the accretion rate in units of $10^{15}\,\rm{g ~ s^{-1}}$, and $B\rm{_{12}}$ is the dipole magnetic field of the pulsar in units of $10^{12}$\,G. For the accretion rates given in Table  \ref{T_phys_param}  and a magnetic field of $\sim10^{12}$\,G, the above formula gives accretion column radii in excellent agreement with those estimated from the spectral fits.

 Key for the formation of the part of the spectrum extending above 10\,keV is, as discussed in \S3.1, a transverse optical depth of at least $\sim1$. 
 From the continuity equation we have that the electron density $n_{e}(z)$ at a given distance $z$ from the center of the neutron star is (c.f. Eq. 43 of F12):

\begin{equation}
  n_{e}(z) =  \frac{ \dot{M}}{v(z) \pi r_{0}^{2} m_{p}}
\end{equation}
 where $\dot{M}$ is the mass infall rate,  $v(z)$ is the velocity of the accretion flow at distance from the neutron star center $z$,  $r_{0}$ is the radius of the accretion column, and  $m_{p}$ is the proton mass.   

 Using the velocity profile given in Eq. \ref{Eq:vel}, the above equation becomes
   
\begin{equation}  
n_{e}(z) =  \frac{ \dot{M}}{ \pi c  m_{p} r_{0}^{2}\beta_{0}} (\frac{z}{R})^{-\eta}
\end{equation}

where $c$ is the speed of light, $\beta\rm{_{0}}$ is the velocity (in terms of the speed of light) at the surface of the neutron star (distance from the center of the neutron star $R$), and $\eta$ is the exponent  of the velocity profile (in our case fixed to $0.5$).  

  Then  the transverse optical depth at distance $z$ is 
\begin{equation}
  \tau_{\perp} = n_{e} r_{0} \sigma_{T} = \frac{L R}{ \pi c m_{p} G M r_{0} \beta_{0}} (\frac{z}{R})^{-\eta}
\end{equation}

  Based on the best-fit parameters given in Table \ref{T_sp_fits}, we find that the transverse optical depths at the base of the accretion column are $\sim1$ (Table \ref{T_phys_param}). Kylafis \etal (2014) demonstrated that  even in such relatively low optical depths, thermal and bulk-motion Comptonization can produce a hard spectrum with a photon index $\Gamma \sim1-1.5$ extending up to $\sim200$\,keV. 
  On the other hand, the best-fit Thompson vertical optical depth along the accretion flow is typically $\tau\sim100-400$.

 \subsection{Limitations of the model}

Although the accretion column model provides good fits to the spectra of the 
5 AXPs/SGRs studied in this work, some of the model parameters take unphysical 
values. The two most extreme examples are the 
 magnitude of the velocity, in units of the velocity of light, that the 
gas in the accretion flow would aquire at the neutron star surface
($\beta_{0}$), which reaches  unity if left as a free parameter, 
and the electron temperature in the accretion flow ($kT_{e}$), which is much 
lower than 1\,keV (Tables \ref{T_sp_fits}, \ref{T:4U0142_fit}). 
This indicates that the spectral model produces the high-energy photons 
only through the bulk-motion Comptonization mechanism, while the contribution 
of thermal Comptonization is negligible, even when the energy available 
through the bulk-motion Comptonization mechanism is not sufficient to 
produce the hard X-ray photons (as in the case when $\beta_{0}$ was fixed 
to a lower value; see \S3).  This behavior could be the result of the 
following factors: 

(a) the seed photon spectrum assumed in the F12 model is a blackbody produced 
at the base of the accretion column, and it ignores the contribution of the 
bremsstrahlung radiation produced by the electrons  below the shock.  In
fact, the F12 model does not take into account at all the flow below the 
radiative shock, but instead assumes that the accretion flow (above the shock)
has a uniform temperature $T_e$. For consistency, the model ought to take into
account the bremsstrahlung radiation that is produced 
from the electrons in the accretion flow (above the shock), where the same 
electrons upscatter the produced photons, thus self-consistently  linking the 
observed spectrum below $\sim20$\,keV with the hard X-ray spectrum. 
Since a bremsstrahlung component would 
contribute in terms of a harder seed spectrum, the required contribution 
of bulk-motion Comptonization would be reduced resulting in more realistic 
values for the terminal velocity of the accretion flow. 

(b) the electron temperature is decoupled from the parameters of the accretion 
flow. In a more realistic scenario one would expect that the same electrons 
that are participating in the bulk-motion above the shock are also responsible 
for the thermal Comptonization below the shock (Lyubarskii \& Sunyaev 1982; 
Kylafis \etal 2014), resulting in much higher electron temperatures and 
harder Comptonized spectra, reducing the need to interpret the entire hard 
spectrum through bulk-motion Comptonization. 

(c) the mass accretion rate estimated through Eq. (3) assumes isotropic 
emission, which obviously would be inappropriate if beaming effects are 
coming into play. In their study of the energy dependent pulse-profiles of 
4U  0142+61, the brightest AXP, Tr{\"u}mper \etal (2013) estimate a luminosity 
for the fan beam of $1.8\times10^{35}$\ergs, and  for the  polar beam 
$0.3\times10^{35}$\ergs, resulting in a total luminosity of 
$2.1\times10^{35}$\ergs in the 0.8-160\,keV energy range.  This is 
$\sim40$\% lower than the luminosity estimated from Eq. (3), indicating 
that the neglect of  beaming effects may significantly overestimate the 
accretion rate, and subsequently any parameters depending on them (e.g. 
optical depth).     

Nonetheless, this model is the best available spectral model for the accretion 
flow in an accreting pulsar, and it provides a good representation of the 
observed spectra. Improved models addressing the above limitations would help 
to obtain more precise information for the parameters of the accretion flow, 
but they should not change the general picture presented here.

\subsection{AXPs and SGRs as accretion-powered pulsars}

 The debate regarding the interpretation of the quiescent emission of AXPs and SGRs as magnetars would largely benefit from the measurement of their dipole megnetic field strengths from cyclotron lines. For example,  more sensitive observations of 1E 1841-04, which show evidence for an absorption feature at the $3\sigma$ level (An \etal 2013),  can give insights into their nature by comparing their spectral and timing properties with those expected from different models. In addition, recently, Tiengo \etal (2013) reported an inferred magnetic field  of  $>2\times10^{14}$\,G  in SGR 0418+5729 assuming a proton cyclotron line, which is interpreted as a multipole magnetic field; on the other hand, if an electron cyclotron line is assumed, the strength of the magnetic field is closer to $\sim10^{11}$\,G.
 The fact that the broad-band X-ray spectra of five AXPs/SGRs with hard X-ray emission can be well reproduced by an accretion flow model onto a pulsar, and the resulting parameters are consistent with those expected from the fall-back disk model, despite the limitations of the spectral models,   gives further support to the fall-back disk model.  Furthermore,  fits with the original version of the BW07 model, which includes bremsstrahlung seed photons (but has strong correlations and degeneracies between the model parameters) also provide an equally good description to the spectral  data.

Tr{\"u}mper \etal (2013) find that the energy-dependent pulse profiles of the brightest AXP, 4U  0142+61, can be explained by the emission of a fan beam originating in the accretion shock and a polar beam from the polar cap. The latter is produced by the illumination of the polar cap by the fan beam, which is subject to gravitational bending towards the neutron-star surface. This model allows us to determine the inclination of the spin axis with respect to the line-of-sight ($i=60$\deg), the angle between the magnetic and spin axes ($\alpha=30$\deg), and the height of the accretion shock from the neutron star surface ($H=2$\,km, corresponding to $z=14.5$\,km from the neutron star center).
 The combination of the constraints from the timing analysis, with the results from the energy spectra presented in this work, provide support to the accretion model for AXPs/SGRs from two independent, but complementary perspectives.   
  Development of more sophisticated spectral models for the accretion flow as well as modeling of the phase resolved spectra (e.g. Tr\"{u}mper \etal 2013) for all sources with good quality hard X-ray spectra will allow us to further test the fall-back disk model for this class of objects and better constrain the parameters of the accretion flow.

\section{Conclusions}

 Our main conclusions are enumerated below.

\begin{enumerate}
\item{ We presented a systematic analysis of the X-ray spectra of 5 AXPs/SGRs. We fitted their \suzaku\ and \int\ data with the spectral model of F12 for the radiation from material infaling onto a pulsar through an  accretion column. We found that this model gives good fits to the observed broad band (0.5-200 keV) spectra, supporting the fall-back disk model.}

\item{ We measured accretion rates in the $\sim10^{15}\rm{g \, s^{-1}}$ range, also in agreement with the calculations of Ertan \etal (2009).}

\item{ We found that the Thompson optical depth along the accretion column is typically $\sim100-400$, while the transverse optical depth is $\sim1.0$.}

\item{ The radius of the accretion column is $\sim200$\,m, consistent with the estimation of Kylafis \etal (2014) for the fall-back disk  model. }

\item{These results are in very good agreement with the independent constraints on the X-ray beaming accretion flow geometry from the analysis of the energy-dependent pulse profiles (Tr{\"u}mper \etal 2013).}

\item{  In general, these results show that the observed broad-band X-ray spectra of AXPs and SGRs showing significant X-ray emission above 10\,keV are consistent with the fall-back disk model. }
 
\end{enumerate}

\section*{Acknowledgments}

AZ acknowledges funding from the European Research Council under the European Union's Seventh Framework Programme (FP/2007-2013) / ERC Grant Agreement n. 617001.
NDK acknowledges partial support by the ``RoboPol'' project, which is implemented under the ``ARISTEIA'' Action of the ``OPERATIONAL PROGRAM EDUCATION AND LIFELONG LEARNING'' and is co-funded by the European Social Fund (ESF) and National Resources.

{}


\begin{thebibliography}{}

\bibitem[{{Alpar} {et~al}(2001)}]{Alpar2001} Alpar, M. A. 2001, \apj, 554, 1245 

\bibitem[{{Alpar} {et~al}(2011)}]{Alpa2011} Alpar, M. A., Ertan, \"U., \& Caliscan, S. 2011, \apj, 732, L4

\bibitem[{{Aguilera} {et~al}(2008)}]{Aguilera2008} Aguilera, D. N., Pons, J., \& Miralles, J. A. 2008,  \apjl, 673, 167

\bibitem[An et al.(2013)]{2013ApJ...779..163A} An, H., Hasco{\"e}t, R., 
Kaspi, V.~M., et al.\ 2013, \apj, 779, 163 

\bibitem[Arnaud(1996)]{1996ASPC..101...17A} Arnaud, K.~A.\ 1996, 
Astronomical Data Analysis Software and Systems V, 101, 17

\bibitem[{{Basko} \& {Sunyaev}(1976)}]{Basko76} Basko, M. M., \& Sunyaev, R. A. 1976, MNRAS, 175, 395 (BS76)

\bibitem[{Becker \& Wolff(2005)}]{2005ApJ...630..465B} Becker, P.~A., \& Wolff, M.~T.\ 2005, \apj, 630, 465 

\bibitem[{{Becker} \& {Wolff}(2007)}]{BW07} Becker, P. A., \& Wolff, M. T. 2007, \apj, 654, 435 (BW07)

\bibitem[{Beloborodov (2013)}]{2013ApJ...762...13B} Beloborodov, A.~M.\ 2013, 
\apj, 762, 13 

\bibitem[{{\c C}al{\i}{\c s}kan \& Ertan(2012)}]{2012ApJ...758...98C} {\c C}al{\i}{\c s}kan, {\c S}., \& Ertan, {\"U}.\ 2012, \apj, 758, 98 

\bibitem[{{Chatterje} {et~al}(2000)}]{Chatterje2000} Chatterjee, P., Hernquist, L., \& Narayan, R. 2000, \apj, 534, 373 

\bibitem[{Dhillon et al.(2011)}]{2011MNRAS.416L..16D} Dhillon, V.~S., Marsh, 
T.~R., Littlefair, S.~P., et al.\ 2011, \mnras, 416, L16 

\bibitem[{Dhillon et al.(2009)}]{2009MNRAS.394L.112D} Dhillon, V.~S., Marsh, 
T.~R., Littlefair, S.~P., et al.\ 2009, \mnras, 394, L112 

\bibitem[{{Duncan} \& {Thompson}(1992)}]{duncan92} Duncan, R. A., \& Thompson, C. 1992, \apj, 392, 9 

\bibitem[{{den Hartog} {et~al}(2008)}]{den Hartog08} den Hartog, P. R., Kuiper, L., Hermsen, W., et al. 2008, A\&A, 489, 245 

\bibitem[{{Enoto} {et~al}(2010)}]{Enoto10} Enoto, T., Nakazawa, K., Makishima, K. et al. 2010, \apjl, 722, 162

\bibitem[{{Ertan} {et~al}(2007)}]{Ertan07} Ertan, \"{U}., Erkut, M. H., Ek\c si, K. Y., \& Alpar, M. A. 2007, \apj, 657, 441 

\bibitem[{Ertan et al.(2009)}]{2009ApJ...702.1309E} Ertan, {\"U}., Ek{\c s}i, 
K.~Y., Erkut, M.~H., \& Alpar, M.~A.\ 2009, \apj, 702, 1309 

\bibitem[{Farinelli et al.(2012)}]{2012A&A...538A..67F} Farinelli, R., Ceccobello, C., Romano, P., \& Titarchuk, L.\ 2012, \aap, 538, A67 

\bibitem[{Farinelli et al.(2008)}]{2008ApJ...680..602F} Farinelli, R., 
Titarchuk, L., Paizis, A., \& Frontera, F.\ 2008, \apj, 680, 602 

\bibitem[{{Ferrigno} {et~al}(2009)}]{Ferrigno09} Ferrigno, C., Becker, P. A., Segreto, A., Mineo, T., \& Santangelo, A. 2009, \aap, 498, 825

\bibitem[{Foster et al. (2012)}]{}Foster, A. R.; Ji, L.; Smith, R. K.; Brickhouse, N. S., 2012, \apj, 756, 128

\bibitem[{Fukazawa et al.(2009)}]{2009PASJ...61S..17F} Fukazawa, Y., Mizuno, 
T., Watanabe, S., et al.\ 2009, \pasj, 61, 17 

\bibitem[{Guo et al.(2015)}]{2015RAA....15..525G} Guo, Y.-J., Dai, S., Li, 
Z.-S., et al.\ 2015, Research in Astronomy and Astrophysics, 15, 525 

\bibitem[{Hasco{\"e}t et al.(2014)}]{2014ApJ...786L...1H} Hasco{\"e}t, R., 
Beloborodov, A.~M., \& den Hartog, P.~R.\ 2014, \apjl, 786, LL1 

\bibitem[{{Kaplan} {et~al}(2009)}]{Kaplan09} Kaplan, D. L., Chakrabarty, D., Wang, Z., \& Wachter, S. 2009, \apj, 700, 149

\bibitem[{Kern \& Martin(2002)}]{2002Natur.417..527K} Kern, B., \& Martin, C.\ 2002, \nat, 417, 527 

\bibitem[{Kouveliotou et al.(1998)}]{1998Natur.393..235K} Kouveliotou, C., 
Dieters, S., Strohmayer, T., et al.\ 1998, \nat, 393, 235 


\bibitem[{Kokubun et al.(2007)}]{2007PASJ...59S..53K} Kokubun, M., Makishima, K., Takahashi, T., et al.\ 2007, \pasj, 59, 53 

\bibitem[{Koyama et al.(2007)}]{2007PASJ...59S..23K} Koyama, K., Tsunemi, H., Dotani, T., et al.\ 2007, \pasj, 59, 23 

\bibitem[{Krivonos et al.(2007)}]{2007A&A...463..957K} Krivonos, R., Revnivtsev, M., Churazov, E., et al.\ 2007, \aap, 463, 957 

\bibitem[{Kuiper et al.(2004)}]{2004ApJ...613.1173K} Kuiper, L., Hermsen, W., \& Mendez, M.\ 2004, \apj, 613, 1173

\bibitem[{Kuiper et al.(2006)}]{2006ApJ...645..556K} Kuiper, L., Hermsen, W., den Hartog, P.~R., \& Collmar, W.\ 2006, \apj, 645, 556 

\bibitem[{Kylafis et al.(2014)}]{2014A&A...562A..62K} Kylafis, N.~D., Tr{\"u}mper, J.~E., \& Ertan, {\"U}.\ 2014, \aap, 562, AA62 

\bibitem[{Lebrun et al.(2003)}]{2003A&A...411L.141L} Lebrun, F., Leray, J.~P., Lavocat, P., et al.\ 2003, \aap, 411, L141 

\bibitem[{Lyubarskii \& Syunyaev (1982)}]{1982SvAL....8..330L} Lyubarskii, Y.~E., \& Syunyaev, R.~A.\ 1982, Soviet Astronomy Letters, 8, 330 

\bibitem[{Mastichiadis \& Kylafis(1992)}]{1992ApJ...384..136M} Mastichiadis, A., \& Kylafis, N.~D.\ 1992, \apj, 384, 136 

\bibitem[{{Mereghetti} (2008)}]{Mereghetti08} Mereghetti, S. 2008, A\&ARv, 15, 225 

\bibitem[{Moretti et al. (2009)}]{2009A&A...493..501M} Moretti, A., Pagani, C., Cusumano, G., et al.\ 2009, \aap, 493, 501 

\bibitem[{Rea et al.(2010)}]{2010Sci...330..944R} Rea, N., Esposito, P., 
Turolla, R., et al.\ 2010, Science, 330, 944 

\bibitem[{Rea et al.(2014)}]{2014ApJ...781L..17R} Rea, N., Vigan{\`o}, D., 
Israel, G.~L., Pons, J.~A., \& Torres, D.~F.\ 2014, \apjl, 781, LL17 

\bibitem[{Rea et al.(2013)}]{2013ApJ...770...65R} Rea, N., Israel, G.~L., 
Pons, J.~A., et al.\ 2013, \apj, 770, 65 

\bibitem[{Smith et al.(2001)}]{2001ApJ...556L..91S} Smith, R.~K., Brickhouse, 
N.~S., Liedahl, D.~A., \& Raymond, J.~C.\ 2001, \apjl, 556, L91 

\bibitem[{{Thompson} \& {Duncan}(1995)}]{Thompson95} Thompson, C., \& Duncan, R. C. 1995, \mnras, 275, 255  

\bibitem[{Titarchuk et al.(1997)}]{1997ApJ...487..834T} Titarchuk, L., 
Mastichiadis, A., \& Kylafis, N.~D.\ 1997, \apj, 487, 834 

\bibitem[{Tr{\"u}mper et al.(2013)}]{2013ApJ...764...49T} Tr{\"u}mper, J.~E., 
Dennerl, K., Kylafis, N.~D., Ertan, {\"U}., 
\& Zezas, A.\ 2013, \apj, 764, 49 

\bibitem[{Tr{\"u}mper et al. (2010)}]{2010A&A...518A..46T} Tr{\"u}mper, J.~E., Zezas, A., Ertan, {\"U}., \& Kylafis, N.~D.\ 2010, \aap, 518, AA46 

\bibitem[{Ubertini et al.(2003)}]{2003A&A...411L.131U} Ubertini, P., Lebrun, F., Di Cocco, G., et al.\ 2003, \aap, 411, L131 

\bibitem[{{van Paradijs} {et~al}(1995)}]{vanParadijs95} 
van Paradijs, J., Taam, R. E., \& van den Heuvel, E. 1995, A\&A, 299, L41

\bibitem[{Vedrenne et al.(2003)}]{2003A&A...411L..63V} Vedrenne, G., Roques, J.-P., Sch{\"o}nfelder, V., et al.\ 2003, \aap, 411, L63 

\bibitem[{Olausen \& Kaspi(2014)}]{2014ApJS..212....6O} Olausen, S.~A., \& Kaspi, V.~M.\ 2014, \apjs, 212, 6 

\bibitem[{Tiengo et al.(2013)}]{2013Natur.500..312T} Tiengo, A., Esposito, P., Mereghetti, S., et al.\ 2013, \nat, 500, 312 

\bibitem[{Walter et al.(2010)}]{2010int..workE.162W} Walter, R., Rohlfs, R., 
Meharga, M.~T., et al.\ 2010, Eighth Integral Workshop.~The Restless 
Gamma-ray Universe (INTEGRAL 2010), 162 

\bibitem[{{Woods}\& {Thompson}(2006)}]{Woods06} 
Woods,  P. M., \& Thompson, C. 2006, in ``Compact Stellar X-ray Sources'', eds. W.H.G. Lewin and M. van der Klis, Cambridge Univ. Press  (astro-ph/0406133) 

\bibitem[{Zane et al.(2009)}]{2009MNRAS.398.1403Z} Zane, S., Rea, N.,  Turolla, R., \& Nobili, L.\ 2009, \mnras, 398, 1403 
\end{thebibliography}
\end{document}